# Solar axions as an energy source and modulator of the Earth magnetic field

*V.D. Rusov*[1*], *E.P. Linnik*[1], *K. Kudela*[2], *S.Cht. Mavrodiev*[3],
*T.N. Zelentsova*[1], *V.P. Smolyar*[1], *K.K. Merkotan*[1]

[1]Department of Theoretical and Experimental Nuclear Physics,
Odessa National Polytechnic University, 65044 Odessa, Ukraine

[2]Institute of Experimental Physics, SAS, Kosice, Slovakia,

[3]The Institute for Nuclear Research and Nuclear Energy, BAS, 1874 Sofia, Bulgaria

**Abstract**

We show existence of strong negative correlation between the temporal variations of magnetic field toroidal component of the solar tachocline (the bottom of convective zone) and the Earth magnetic field (Y-component). The possibility that hypothetical solar axions, which can transform into photons in external electric or magnetic fields (the inverse Primakoff effect), can be the instrument by which the magnetic field of convective zone of the Sun modulates the magnetic field of the Earth is considered.

We propose the axion mechanism of "solar dynamo–geodynamo" connection, where an energy of axions, which form in the Sun core, is modulated at first by the magnetic field of the solar tachocline zone (due to the inverse coherent Primakoff effect) and after that is absorbed in the liquid core of the Earth under influence of the terrestrial magnetic field, thereby playing the role of an energy source and a modulator of the Earth magnetic field. Within the framework of this mechanism new estimations of the strength of an axion coupling to a photon ($g_{a\gamma} \sim 5 \cdot 10^{-9}$ GeV$^{-1}$) and the axion mass ($m_a \sim 30$ eV) have been obtained.

---

[*] Corresponding author: V.D. Rusov, e-mail siiis@te.net.ua

***Introduction***. It is known that in spite of a long history the nature of the energy source maintaining a convection in the liquid core of the Earth or, more exactly, the mechanism of the magnetohydrodynamic dynamo (MHD) generating the magnetic field of the Earth still has no clear and unambiguous physical interpretation [1-5]. The problem is aggravated because of the fact that none of candidates for an energy source of the Earth magnetic-field [1] (secular cooling due to the heat transfer from the core to the mantle, internal heating by radiogenic isotopes, e.g., $^{40}K$, latent heat due to the inner core solidification, compositional buoyancy due to the ejection of light element at the inner core surface) can not in principle explain one of the most remarkable phenomena in solar-terrestrial physics, which consists in strong (negative) correlation between the temporal variations of magnetic flux in the tachocline zone (the bottom of the Sun convective zone) [6,7] and the Earth magnetic field [8] (Fig. 1).

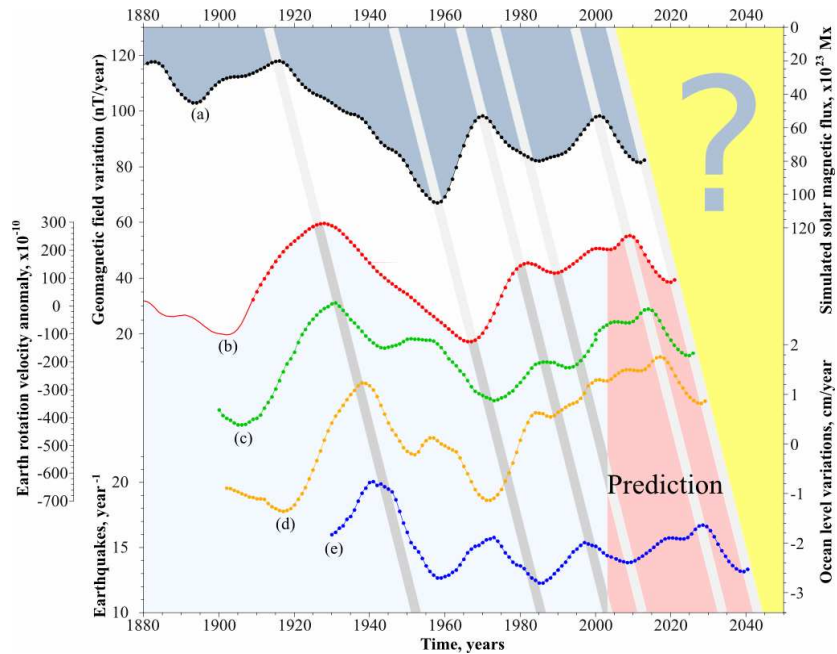

**Fig. 1.** Time evolution (a) the variations of magnetic flux in the bottom (tachocline zone) of the Sun convective zone (see Fig.7f in Ref. [6]), (b) of the geomagnetic field secular variation (*Y*-component, nT/year) [8], (c) the variation of the Earth's rotation velocity [9], (d) the variation of the average ocean level (PDO+AMO, cm/year) [10] and (e) the number of large earthquakes (with the magnitude M≥7) [11] and. All curves are smoothed by the sliding intervals in 5 and 11 years. The pink area is a prediction region. Note: formation of the second peaks on curves (c)-(e) is mainly predetermined by nuclear tests in 1945-1990.

At the same time, supposing that the transversal (radial) surface area of tachocline zone, through which a magnetic flux passes, is constant in the first approximation, we can consider that magnetic flux variations describe also the temporal variations of magnetic field in the tachocline zone of the Sun. In fact, it is no wonder because the strong correlation between the temporal variations of the Earth magnetic field and the North-South asymmetry of solar equatorial rotation [7,12,13], which, indeed, is exactly the physical cause of strong negative correlation of temporal variations of the Earth magnetic field and the magnetic field toroidal component in the tachocline zone of the Sun [7], has been known for a long time. In this sense, it is obvious that a future candidate for an energy source of the Earth magnetic field must play not only the role of a natural trigger of solar-terrestrial connection, but also directly generate the solar-terrestrial magnetic correlation by its own participation.

The fact that the solar-terrestrial magnetic correlation has, undoubtedly, fundamental importance for evolution of all the Earth's geospheres is confirmed by existence of stable and strong correlation between temporal variations of the Earth magnetic field, the Earth angular velocity, the average ocean level and the number of large earthquakes (with the magnitude M≥7), whose generation is apparently predetermined by a common physical cause of unknown nature (see Fig. 1).

In this paper we consider hypothetical particles (solar axions) as the main carriers of the solar-terrestrial connection, which by virtue of the inverse coherent Primakoff effect can transform into photons in external fluctuating electric or magnetic fields [14]. At the same time we ground and develop the axion mechanism of "solar dynamo−geodynamo" connection, where the energy of axions, which form in the Sun core, is modulated at first by the magnetic field of the solar tachocline zone (due to the inverse coherent Primakoff effect) and after that is absorbed in the liquid core of the Earth under the influence of the terrestrial magnetic field, thereby plays the role of an energy source and a modulator for the Earth magnetic field. Justification of axion's mechanism of the "solar dynamo−geodynamo" connection is the goal of this paper.

***Implication from "axion helioscope" technique.*** As it seen from the Earth, the most important astrophysical source for axions is the core of the Sun. There, pseudoscalar particles like axions would be continuously produced in the fluctuating electric and magnetic fields of the plasma via their coupling to two photons. After production the axions would freely stream out of the Sun without any further interaction[*]. Resulting differential solar axion flux on the Earth would be [15, 16]

$$\frac{d\Phi_a}{dE} = 6.02 \cdot 10^{10} g_{10}^2 E^{2.481} \exp\left(-\frac{E}{1.205}\right) \quad cm^{-2} s^{-1} keV^{-1}. \qquad (1)$$

The spectral energy of the axions (1) follows a thermal energy distribution between 1 and 100 keV, which peaks at ≈ 3 keV and the average energy $<E_a>$=4.2 keV. To be able to compare the expected axion flux in a specific energy range with available data, the spectrum given by (1) is integrated over the energy range of 1 to 100 keV find the solar axion flux at Earth to be

$$\Phi_a \approx 3.75 \cdot 10^{11} g_{10}^2 \quad cm^{-2} s^{-1}. \tag{2}$$

---

[*] Given that the axion mean free path in the Sun is ≈ $g_{10}^{-2} \cdot 6 \cdot 10^{19} km$ for 4 keV axions [15]. Here we define $g_{10}=g_{a\gamma\gamma}/10^{-10}$ GeV$^{-1}$, where $g_{a\gamma\gamma}$ is axion to photon coupling constant.

In case of the coherent Primakoff effect the number of photons leaving the magnetic field towards the detector is determined by the probability $P_{a \to \gamma}$ that an axion converts back to a "observable" photon inside the magnetic field [17]

$$P_{a \to \gamma} = \left(\frac{B g_{a\gamma\gamma}}{2}\right)^2 \frac{1}{q^2 + \Gamma^2/4} \left[1 + e^{-\Gamma L} - 2e^{-\Gamma L/2} \cos(qL)\right], \tag{3}$$

where $B$ is strength of the transverse magnetic along the axion path, $L$ is the path length traveled by the axion in magnetic region, $l=2\pi/q$ is the oscillation length, $\Gamma=\lambda^{-1}$ is the absorption coefficient for the X-rays in the medium, $\lambda$ is the absorption length for the X-rays in the medium and the longitudinal momentum $q$ difference between the axion and an X-rays energy $E_\gamma = E_a$ is

$$q = \frac{|m_\gamma^2 - m_a^2|}{2E_a} \tag{4}$$

with the effective photon mass

$$m_\gamma \cong \sqrt{\frac{4\pi \alpha n_e}{m_e}} = 28.9 \sqrt{\frac{Z}{A} \rho}, \tag{5}$$

where $m_a$ is the axion mass, $\alpha$ is fine-structure const, $n_e$ is the number of electrons in the medium, $m_e$ is the electron mass, $Z_{eff}$ is the free density electrons of the medium, $A$ is atomic mass of the medium and its density $\rho$ in g/cm$^3$. For a specific pressure and temperature, the coherence is restored for a narrow axion mass window, for which the effective mass of the photon matches that of the axion such that

$$qL < \pi \Rightarrow \sqrt{m_\gamma^2 - \frac{2\pi E_a}{L}} < m_a < \sqrt{m_\gamma^2 + \frac{2\pi E_a}{L}}. \tag{6}$$

***Axion conversion in the Sun magnetic field and the plasma mass of photon***. Let us consider modulation of an axion flux emerging from the Sun core on passing through the solar tachocline region (ST) located in the base of convective zone of Sun (Fig. 2). As is known [18],

The thickness of ST, where the toroidal magnetic field $B\sim20\div50$ T dominates, attains $L_{ST}$ $\sim0.05R_S \approx 3.5\cdot10^4$ km (where $R_S$ is the Sun radius). At the same time the values of pressure, temperature and density for the ST are $P_{ST} \sim 8\cdot10^{12}$ Pa, $T_{ST} \sim 2\cdot10^6$ K and $\rho \sim 0.2$ g·cm$^{-3}$, respectively.

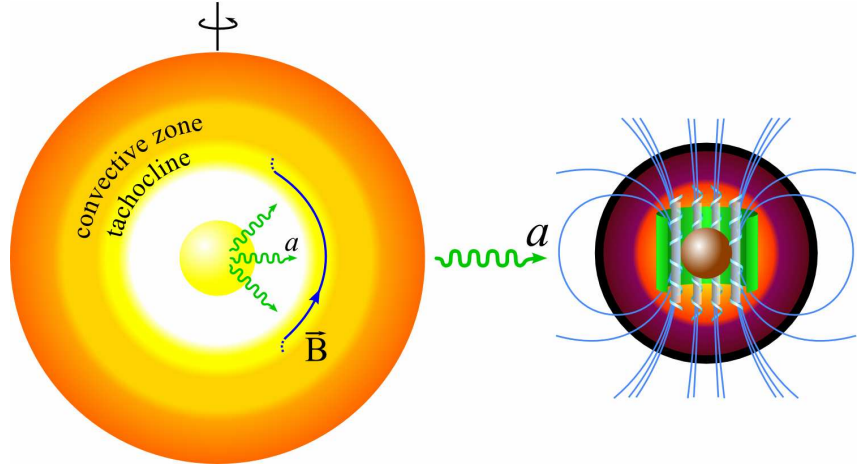

**Fig.2.** Schematic representation of the solar tachocline zone and Earth's liquid outer core (red region) in whose magnetic fields the conversion of axions into γ-quanta can take place with certain probability (see Eq. (3)). **Note:** In the conventional concept, the molten iron of liquid phase of Earth's core circulates along a spiraling in columns aligned in the north-south direction, generating electrical currents that set up the dipolar magnetic field. The concentration of fieldlines into anticyclonic vortices (rotating in the same as air around a region of high pressure) has been thought to explain the intense magnetic lobes found in Earth's field at the top of the core. (The region contained by Earth's inner core and green cylinder is characterized by strong magnetic field.)

To estimate the plasma mass of photon $m_\gamma$ in the hydrogen-helium medium of ST it is possible, without loss of generality, to use the modified Eq. (5) in the form [15]

$$m_\gamma \cong 28.9\sqrt{\frac{Z}{A}\rho} = \sqrt{0.02\frac{P_{ST}(mbar)}{T_{ST}(K)}} \sim 30 \quad eV, \qquad (7)$$

where we use the corresponding parameters $T_{ST}$ and $P_{ST}$ for the hydrogen-helium medium of ST.

Now we make an important assumption that the axion mass is equal to the plasma mass of photon, i.e., $m_a\sim30$ eV. It is obvious, that by virtue of Eqs. (4) and (7) $q\to0$, whence it follows that the oscillation length $l$ becomes infinite quantity, i.e., $l=2\pi/q \to \infty$. However, taking into account that in this case the absorption length $\lambda$ is about 0.1 m [19], we have $\Gamma L_{ST} \to \infty$. This means that according to Eq. (3) the intensity of expected conversion of axions into γ-quanta is practically equal to zero in this case.

At the same time, there is a reason to believe (see Ref. [20] and Refs. therein) that in fact the conversion of axions into γ-quanta, apparently, takes place and, strangely enough, this process goes on quite effectively. For example, the reconstructed solar photon spectrum below 10 keV from the Active Sun (Fig. 3) well-describable by the sum of secondary Compton's spectra obtained, for example, by the simulation of passage of γ–quanta (regenerated from solar axion spectrum in the tachocline zone of the Sun (Fig.3, dashed line)) through the areas of solar photosphere of different thickness but equal density (Fig.3, layers with the thickness of 64 g/cm$^2$, 16 g/cm$^2$ and 2 g/cm$^2$).

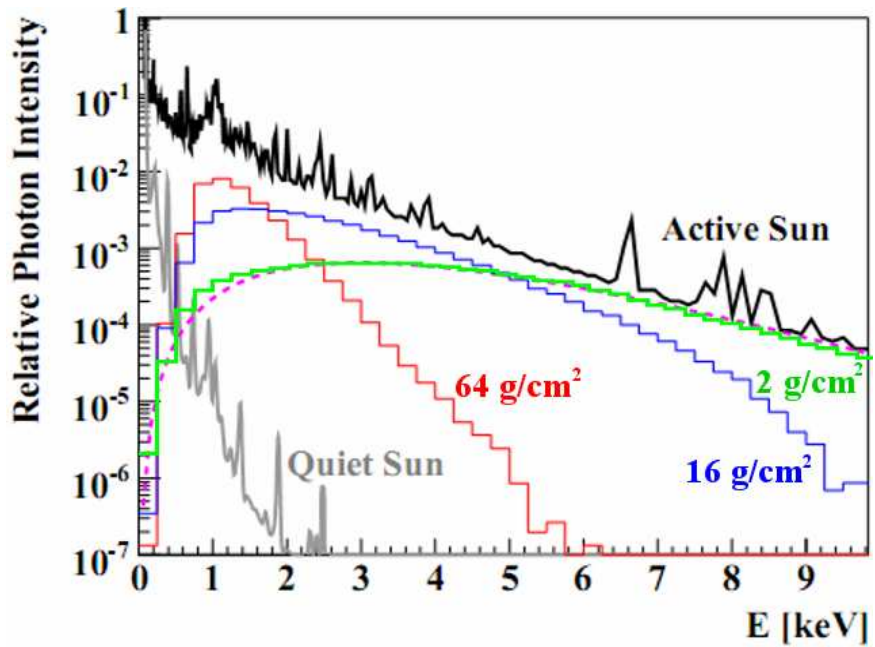

**Fig.3.** Reconstructed solar photon spectrum below 10 keV from the Active (flaring) Sun (black line) from accumulated observations (adapted from [20]). The dashed line is the converted solar axion spectrum. Three degraded spectra to multiple Compton scattering are also shown for column densities above the initial conversion place of 64 g/cm$^2$, 16 g/cm$^2$ [20] and 2 g/cm$^2$ (present paper). Note that the Geant4 code photon threshold is at 1 keV and therefore the turndown around ~1 keV is an artefact.

In other words, despite the fact that the coherent axion-photon conversion by the Primakoff effect is impossible due to the small absorption length for γ–quanta ($l \ll 1$) in the medium (see Eq.(3) $\Gamma = 1/\lambda \to \infty$), there is a good agreement between the relative theoretical γ–quantum spectra generated by solar axions and experimental photon energy spectra detected close to the Sun surface in the period of its active phase (see Fig. 3). At the same time it is necessary to note that attempts to match the absolute values of these spectra are not succeeded until now [20].

To overcome the problem of the small absorption length for $\gamma$-quanta and to reach a resonance in Eq. (3) it is necessary that a refractive gas, in which the axion-photon oscillation is researched, would have zero index of refraction [21]. It appears that to satisfy this condition it is not required to use so-called metamaterials [22] with the negative permittivity ($\varepsilon$) and magnetic permeability ($\mu$) or results of the Pendry superlens theory [23], which are not practically realized in nature. We can use the results of work [24], where the possibility of channeling of electromagnetic radiation in a microwave range in a multi-layered metal-dielectric structure with the different (it is very important!) thicknesses of alternating layers is theoretically and experimentally shown. And one of alternating layers must have the negative permittivity $\varepsilon$, that is typical not only for a metal but also for plasma [22].

Taking into account that a plasma medium is successfully simulated by an analogical multi-layer metal-dielectric structure [25], we can assume that a plasma medium can have zero index of refraction under certain conditions, for example, in the tachocline zone of the Sun. This means, in its turn, that the absorption length $\lambda$ for photons in such a medium will become considerably greater than the thickness of tachocline zone, i.e., $\lambda \gg L_{ST}$. At the same time, it is obvious that $\Gamma = \lambda^{-1} \to 0$, whence it follows the necessary condition **$\Gamma L_{ST} \to 0$.**

In this case the probability (3) that an axion converts back to a "observable" photon inside the magnetic field can be represented by the following simple form

$$P_{a\gamma} \cong \left( \frac{g_{a\gamma} B L}{2} \right)^2, \tag{8}$$

Within the framework of the present paper we take rather conservative values for the magnetic field $B=20$ T and the thickness of solar tachocline zone $L_{ST}=2 \cdot 10^7$ m. We have chosen the effective thickness $L_{ST}$ proceeding from the sufficient degree of uniformity of solar tachocline zone over its temperature, pressure and density. Then using Eqs. (2), (8) and the parameters of ST magnetic field, it is possible to write down the expression for the solar axion flux at the Earth as

$$\Phi_a(L_{Earth}) \cong \Phi_a \cdot P_{a \to \gamma} = \Phi_a \cdot \left( \frac{g_{a\gamma}}{5 \cdot 10^{-9} GeV^{-1}} \right)^2 \left( \frac{B}{20T} \right)^2 \left( \frac{L}{2 \cdot 10^7 m} \right)^2, \tag{9}$$

where $L_{Earth}$ is the distance from the Earth to the Sun.

In normalizing the expression (9) the probability of the total conversion of axions into photons is assumed to be equal to unit at given parameters of ST magnetic field, i.e., $P_{a\to\gamma}(B_{ST}=20$ T, $L_{ST}=2\cdot10^7$ m) =1. Note that the validity of "handmade", at first sight, choice of the strength of an axion–photon coupling ($g_{a\gamma}\sim 5\cdot10^{-9}$ GeV$^{-1}$) and normalizing condition (9) will be given below. So, on the one hand, the value of strength of an axion coupling to a photon is specified

$$g_{a\gamma} \sim 5\cdot 10^{-9} \quad GeV^{-1}, \tag{10}$$

and, on the other hand, the following variations of axion intensity on the Earth surface (see Eq. (2)) will correspond to the variations of ST magnetic activity:

$$\Phi_a(L_{Earth}) \approx 9.375\cdot 10^{14}\cdot \begin{cases} 0 & at \quad B_{CZS}^{max}=20T, \\ 1.0 & at \quad B_{CZS}^{min}=5T, \end{cases} cm^{-2}s^{-1}. \tag{11}$$

It is interesting to note that at the given value of the strength of an axion coupling to a photon (9) the estimation of the axion-converted photon luminosity $\Lambda_\gamma^{Sun}$ at an active phase of the Sun

$$\Lambda_\gamma^{Sun} \cong \Phi_a(E_{Earth})\frac{r_{SE}^2}{r_S^2}\langle E_a\rangle \sim 10^{11} \quad erg/cm^2 s \tag{12}$$

coincides an order of magnitude with the experimentally measured radioactive solar surface brightness [18]. Here $r_S=6.960\cdot 10^{10}$ cm is the radius of the Sun, $r_{SE}=1.496\cdot 10^{13}$ cm is the distance from the Earth to the Sun, $\langle E_a\rangle=4.2$ keV is the axion average energy.

Returning to the validation of normalizing condition (9), we would like to pay attention that in our case the expression (12) is just the selection criterion for the value of strength of an axion coupling to a photon (10).

*Axion conversion in the Earth magnetic field and the plasma mass of photon*. Now let's consider the solar axion intensity modulation by the Earth's magnetic field, when axions pass through the convective zone of the Earth (CZE). As is known, the value of the Earth magnetic field, which prevails near magnetic toroidal tubes in the liquid core of the Earth, is no greater than $B\sim 1.5$ T and it drops as $1/r^3$ [26]. However, in the region bounded by the solid core surface and directly by tubes (Fig. 2) we may assume $B\sim 1.5$ T. At the same time the values of pressure, temperature and density typical for this part of the convective zone are $P_{CZE}\sim 2\cdot 10^{11}$ Pa, $T_{CZE}\sim 4\cdot 10^4$ K and $\rho\sim 11$ g/cm$^3$, respectively. Given the molten iron core, we will treat it as a plasma composed of a degenerate gas of free electrons, moving in the background of Fe ions with effective charge $Z_{eff} \approx 5.4$ [27]. We are also ignoring the contribution of other trace elements, such as nickel, which have more or less the same properties as iron, for our purposes.

To estimate the plasma mass of photon $m_\gamma$ in nickel-iron plasma we use Eq. (7) *again*. Calculations fulfilled for the both modifications of Eq. (7) show that the plasma mass of photon $m_\gamma$ in nickel-iron plasma with the specified parameters for the liquid core of the Earth is

$$m_\gamma \cong 28.9\sqrt{\frac{Z_{eff}}{A_{Fe}}\rho} = \sqrt{0.02\frac{P_{CZE}(mbar)}{T_{CZE}(K)}} \sim 30 \quad eV. \qquad (13)$$

As in our case $m_a \sim 30$ eV, we have $q \to 0$ by virtue of Eqs. (4) and (13). Whence it follows that the oscillation length $l$ becomes infinite quantity, i.e., $l = 2\pi/q \to \infty$. Taking into account that the absorption length $\lambda$ is $\sim 10^{-7}$ m [18] for this case and the double length of CZE bottom (green area in Fig.2) is about $L_{CZE} \sim 1.2 \cdot 10^6$ m, we obtain $\Gamma L_{CZE} \to \infty$. It is obvious, that this one is similar to already considered situation for the solar tachocline zone, where the intensity of expected conversion of axions into γ-quanta is zero too.

On the other hand, surprising coincidence of the values of plasma mass of photon $m_\gamma$ in the tachline zone of the Sun and near-surface region of the liquid core of the Earth and also the indicated above properties of plasma medium make it possible to suppose that the plasma medium in the liquid core of the Earth has zero index of refraction too. So, if the medium has zero index of refraction, then $\Gamma L_{CZE} \to 0$. Using Eqs. (2), (9) and the parameters of CZE magnetic field ($B=1.5$ T; $L_{CZE} \sim 1.2 \cdot 10^6$ m) we can write down the expression for the axion-converted photons intensity near magnetic toroidal tubes in the liquid core of the Earth in the following form

$$\Phi_\gamma(L_{core}) = \Phi_a(L_{Earth}) \cdot P_{a \to \gamma}(L_{CZS}) \sim 1.89 \cdot 10^{10} \quad cm^{-2}s^{-1}, \qquad (14)$$

Obviously, that in this case the correspondence between the variations of CZS magnetic activity ($B \sim 5.0 \div 20$ T) and the variations of converted photon intensity near magnetic toroidal tubes in the liquid core of the Earth looks like

$$\Phi_\gamma(L_{core}) \sim 1.89 \cdot 10^{10} \cdot \begin{cases} 1 & at \quad B_{CZS}^{min} \leq 5.0T, \\ 0 & at \quad B_{CSS}^{max} = 20T, \end{cases} \quad cm^{-2}s^{-1}. \qquad (15)$$

***Power required to maintain the Earth magnetic field and the strength of an axion–photon coupling.*** As the average energy of axions is $<E_a>=4.2$ keV, the maximum energy-release velocity $W_\gamma$ in cylindrical domain of the radius $R \approx 2000$ km and height $H \approx 6000$ km, which near magnetic toroidal tubes (Fig. 2) is approximately equal to the liquid core diameter, with an allowance for (15) looks like

$$W_\gamma = \Phi_\gamma(L_{Core}) \cdot 2RH \cdot \langle E_a \rangle \sim 3 \quad TW. \tag{16}$$

Analysis of modern model parameters of the thermal state of the Earth's core [1] shows that in spite of known hardships in interpretation of the results of evolutionary geodynamo simulation, such a thermal power (3 TW) is sufficient for generation and maintenance of the Earth magnetic-field [1-3]. B. Buffet considers these hardships as a consequence of ignorance of a physical nature of thermal sources in the Earth's core and summarizes the situation in the following words [1]: "*At present time, buoyancy sources due to latent heat release* [28] *and exclusion of light alloying elements* [29] *from the inner core drive vigorous fluid motions with relatively modest heat flows. It is estimated that the power required to maintain the magnetic field is roughly* 1 *TW* [2]. *This rate of power consumption can be sustained at the present time with a heat flow of* 5 *to* 6 *TW* [4]. *Such values are at the low end of current estimates, indicating that the more than enough power is available to generate the magnetic field*".

Thus, so long as this heat is enough for generation and maintenance of the Earth magnetic field, we can consider that our choice of the strength of an axion-photon coupling $g_{a\gamma} \sim 5 \cdot 10^{-9}$ GeV$^{-1}$ and normalization condition (9) is correct and natural within the framework of the considered mechanism. All this applies in full measure to the value of axion mass $m_a \sim 30$ eV.

*Summary.* We have shown that strong correlation between the temporal variations of magnetic field of the Earth (Y-component) and the magnetic field toroidal component of tachocline zone of the Sun really takes place. Thereupon we have asked ourselves: "May solar axions, which can transform into photons in external electric or magnetic fields (the inverse Primakoff effect), be the instrument by which the magnetic field of the solar tachocline zone modulates the magnetic field of the Earth?"

To answer the question we have proposed the axion mechanism explaining the "solar dynamo-geodynamo" connection, where the total energy of axions, which appear in the Sun core, is initially modulated by the magnetic field of the solar tachocline zone due to the inverse coherent Primakoff effect and after that is absorbed in the Earth liquid core also due to the inverse coherent Primakoff effect. It results in the fact that the variations of axion intensity play a role of an energy source and a modulator of the Earth magnetic field. In other words, the solar axion's mechanism is not only responsible for formation of a thermal energy source in the liquid core of the Earth necessary for generation and maintenance of the Earth magnetic field, but unlike other alternative mechanisms [5] (such as secular cooling due to heat transfer from the core to the mantle, internal heating by radiogenic isotopes (e.g., $^{40}K$), latent heat due to the inner core solidification, compositional buoyancy due to the ejection of light element at the inner core surface) naturally explains the cause of experimentally observed strong negative correlation of the magnetic field of tachocline zone of the Sun and magnetic field of the Earth.

Within the framework of this mechanism new estimations of the strength of an axion coupling to a photon ($g_{a\gamma} \sim 5 \cdot 10^{-9}$ GeV$^{-1}$) and the axion mass ($m_a \sim 30$ eV) have been obtained. It is necessary to note that obtained estimations can't be excluded by the existing experimental data [30-38] because the discussed above effect of solar axion intensity modulation by temporal variations of the toroidal magnetic field of the solar tachocline zone was not taken into account in these observations.

Finally, it is possible to conclude that within the framework of axion-solar-terrestrial hypothesis temporal variations of such fundamental geophysical parameters of the Earth as the magnetic field, angular velocity, average ocean level and the number of large earthquakes (with the magnitude M≥7) have the same cause – the temporal variations of solar axion intensity in the Sun magnetic field. And, as follows from Fig. 1, each of these parameters is characterized by a certain time lag relative to the primary parameter predetermined by the axion mechanism. Obviously, such a delay effect makes it possible to predict reliably the behavior of variations of the mentioned parameters by experimental observations of temporal variations of the toroidal magnetic-field of the Sun convective zone or, in the last resort, of the Earth magnetic field.